\documentclass[pra,twocolumn,showpacs,letterpaper,showpacs,superscriptaddress]{revtex4-1}
\usepackage{graphicx,amsmath,amssymb,amsfonts,latexsym,color,dcolumn,bm}
\usepackage{braket}
\usepackage{makeidx}
\usepackage{multirow}
\usepackage[dvipsnames,svgnames,table]{xcolor}
\usepackage{graphicx}
\usepackage{epstopdf}
\usepackage{ulem}

\begin{document}
\newcommand{\be}{\begin{equation}}
\newcommand{\ee}{\end{equation}}
\newcommand{\bea}{\begin{eqnarray}}
\newcommand{\eea}{\end{eqnarray}}
\newcommand{\ad}{a^{\dag}}
\newcommand{\la}{\langle}
\newcommand{\ra}{\rangle}
\newcommand{\om}{\omega}
\newcommand{\Ep}{E^{(+)}}
\newcommand{\Em}{E^{(-)}}
\newcommand{\pa}{\partial}

\title{Complementarity in Biphoton Generation with Stimulated or Induced Coherence}
\author{A. Heuer}
\affiliation{University of Potsdam, Institute of Physics and Astronomy, Carl-Liebknecht Stra\ss e 24-25,
D-14476 Potsdam}
\author{R. Menzel}
\affiliation{University of Potsdam, Institute of Physics and Astronomy, Carl-Liebknecht Stra\ss e 24-25,
D-14476 Potsdam}
\author{P.W. Milonni}
\affiliation{Theoretical Division, Los Alamos National Laboratory, Los Alamos, New Mexico 87545 USA} 
\affiliation{Department of Physics and Astronomy, University of Rochester, Rochester, NY 14627, USA} 
\date{}

\begin{abstract}

Coherence can be induced or stimulated in parametric down-conversion using two or three crystals when, for example, the idler modes of the crystals are aligned. Previous experiments with induced coherence [Phys. Rev. Lett. {\bf 114}, 053601 (2015)] focused on which-path information and the role of vacuum fields in realizing complementarity via reduced visibility in single-photon interference. Here we describe experiments comparing induced and stimulated coherence. Different single-photon interference experiments were performed by blocking one of the pump beams in a three-crystal setup. Each counted photon is emitted from one of two crystals and which-way information may or not be available, depending on the setup. Distinctly different results are obtained in the induced and stimulated cases, especially when a variable transmission filter is inserted between the crystals. A simplified theoretical model accounts for all the experimental results and is also used to address the question of whether the phases of the signal and idler fields in parametric down-conversion are correlated.
\end{abstract}


\maketitle

\section{Introduction}

Complementarity has been recognized as one of the cornerstones of quantum physics ever since it was expounded by Bohr \cite{Bohr28}. It is closely connected to the theory of measurement and in particular to notions of distinguishability. In quantum optics light is detected as photon (or photoelectron) clicks, subject of course to restrictions regarding spatial, spectral, temporal and polarization selectivity. The measured intensity pattern of bright light reflects the statistical distribution of these clicks. The visibility $V$ of the interference fringes is diminished when there is ``which-path" knowledge or distinguishability regarding different photon pathways to the detector. Complementarity in quantum optics therefore relates to the question of how distinguishable are the photon pathways. The visibility $V$ is limited by the relation ${K^2+V^2 \leq 1}$, where $K$ characterizes the distinguishability or which-path ``knowledge." \cite{Jaeger95}.\

It is of course of interest to investigate the detailed physical origin of this complementarity, and the limitation on visibility imposed by which-path information, in specific experiments. In a previous investigation \cite{heuer14-1} it was shown that the randomness of the vacuum fields in spontaneous parametric down-conversion (SPDC) with induced coherence \cite{Zou91,Wang91} limits the visibility of single-photon interference fringes when which-path knowledge is available. In that work there was induced coherence between two SPDC crystals, and a third crystal provided which-path information for the signal photons via the detection of the reference idler photons, and vice versa. Although the signal modes for the two crystals were perfectly matched 
geometrically, their temporal phases could be regarded as completely incoherent as a consequence of the vacuum fields taking part in the spontaneous down-conversion. In other words, limitations on the fringe visibility could be attributed physically to the vacuum fields.

To develop further insights into this complementarity and the role of the vacuum fields in SPDC, we have performed experiments in which the quantum fluctuations of the vacuum fields can be ``over-written" by stimulating coherence in two or three down-conversion crystals. In these experiments the generation of the idler field, for example, is stimulated in the nonlinear crystals by an external laser field. If the laser wavelength lies within the emission band of the spontaneously generated idler photons, and the laser intensity is high enough, the stimulated biphoton generation will be much stronger than the spontaneous biphoton generation \cite{Ou90}, and high visibilities are the consequence. But then the question of which-path information arises once again.

We modified the experiments reported earlier \cite{heuer14-1} by applying a He-Ne laser to stimulate the biphoton production in the SPDC crystals. The He-Ne laser mode was perfectly matched geometrically to the modes of the idler photons of the crystals. With the coherent laser light it was possible to over-ride almost completely the effects of the vacuum fields taking part in SPDC, and visibilities of the single-photon interference patterns were always above 90\%. 

In one of these experiments only two separated SPDC crystals were pumped and seeded in a parallel configuration and not otherwise connected in any way. As a consequence of phase memory \cite{heuer14-2} we obtained very high visibility for both the signal and idler single-photon fields emitted from one of the two crystals, implying low which-path knowledge. In this stimulated coherence experiment complementarity is realized up to the theoretical limit specified by ${K^2+V^2 \leq 1}$. Depending on the details of the experimental setup, the visibility may be reduced by unavoidable spontaneous emission of the biphotons, which again involves the vacuum field randomness as described earlier \cite{heuer14-1}, or the which-path information may be reduced by the stimulating fields.\

A difference between stimulated and induced coherence is also observed when a transmission filter is inserted between the two crystals. In this case the effect on stimulated coherence results simply from the attenuation of the
stimulating field, whereas the effect on induced coherence cannot be viewed as simply a result of vacuum-field attenuation, which would violate the fundamental commutation relation between photon annihilation and creation operators. \ 

In summary, the experiments described here with stimulated and induced coherence allow for a detailed analysis of the physical background of complementarity in this area of quantum optics, and shed further light on the role of the vacuum fields taking part in SPDC. In the following two sections we describe our experiments and present a simplified theoretical model that accounts for the results of these experiments. In Section \ref{sec:phase} we address the question of whether we can regard the phases of the signal and idler fields of a biphoton pair as being correlated. In Section \ref{sec:conclusions} we summarize and briefly discuss our conclusions.

\section{Experimental}
The general experimental setup consists of three BBO crystals for biphoton generation (Fig. 1). All three crystals had a length of 4 mm and were cut for type I phase matching. With angular alignment and spectral filters wavelengths of 808 nm for the signal photons and 633 nm for the idler photons were chosen. The three crystals were arranged  in such a way that the idler channel i1 of crystal BBO1 was matched to the idler channel i2 of crystal BBO2, and the signal channel s1 of BBO1 was matched to the signal channel s3 of BBO3. The three pump beams were obtained from a single laser (Genesis 355, Coherent Inc.) which emitted an almost diffraction-limited cw field at 355 nm. The laser power of 10 mW was split almost equally among the three pump beams  for the three crystals. The stimulating 632.8-nm He-Ne laser operated single-mode with a maximum power of 5 mW, and was attenuated to about 50 $\mu$W. To stimulate the parametric process in all three crystals simultaneously, the He-Ne laser beam was also split.  The two beams were aligned to propagate along paths essentially indistinguishable from the idler paths i1,i2, and i3, respectively (Fig. 1). The signal fields in modes s1, s2 and s3 should then exhibit single-photon, first-order interference at detector A. The phase between these two modes can be varied by moving the $\cong$ 100\%-reflecting mirror (phase 1). Different experiments can be performed by blocking one of the three pump beams.

\begin{figure}[t]
\includegraphics[width=8cm]{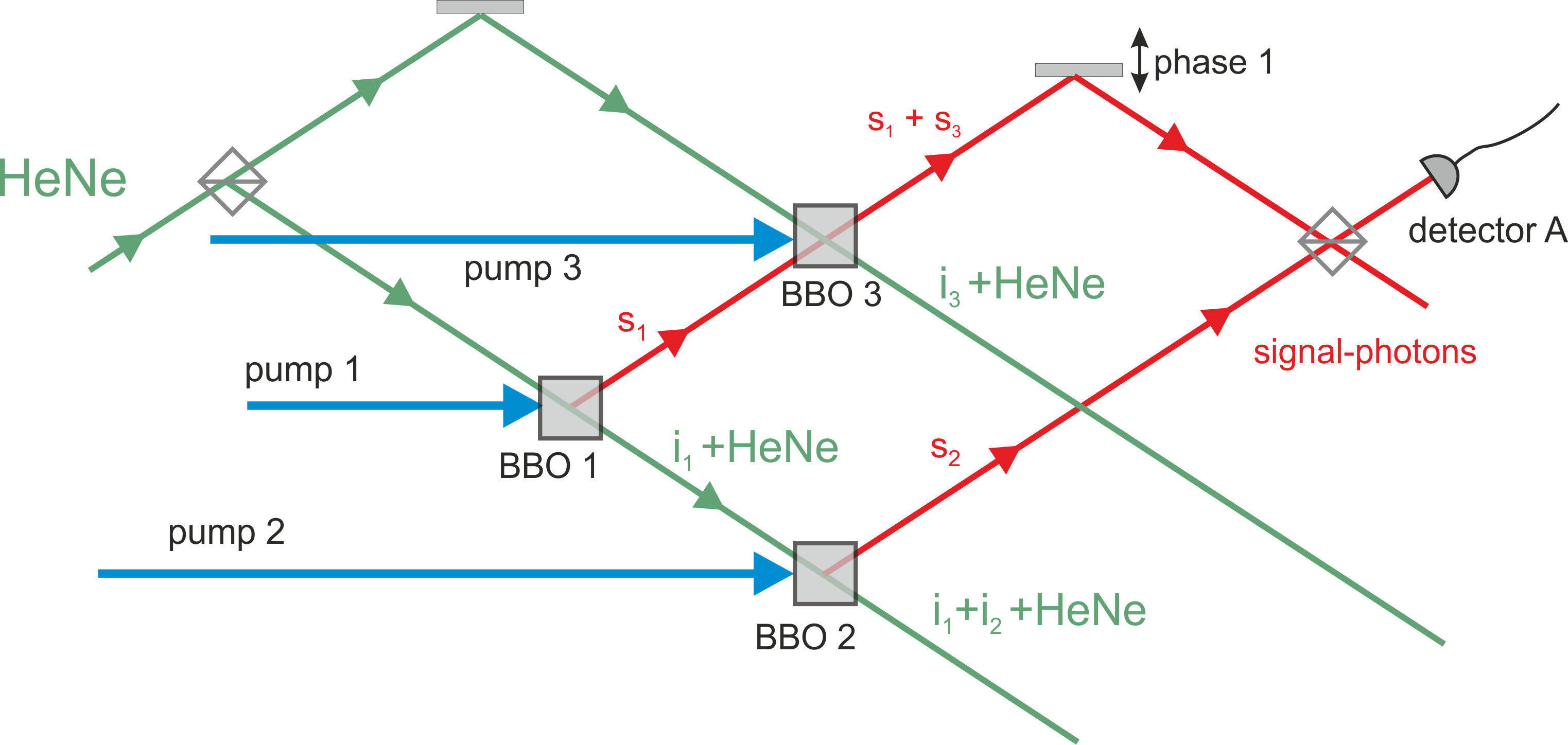}
\caption{Experimental setup for the measurement of single-photon signal interference at detector A. All three crystals BBO1, BBO2 and BBO3 are synchronously pumped in a cascade arrangement and a He-Ne laser stimulates the emission of idler fields from the three crystals.}
\label{Fig1}
\end{figure}

\subsection{BBO1 and BBO2 pumped}
\begin{figure}[t]
\includegraphics[width=7cm]{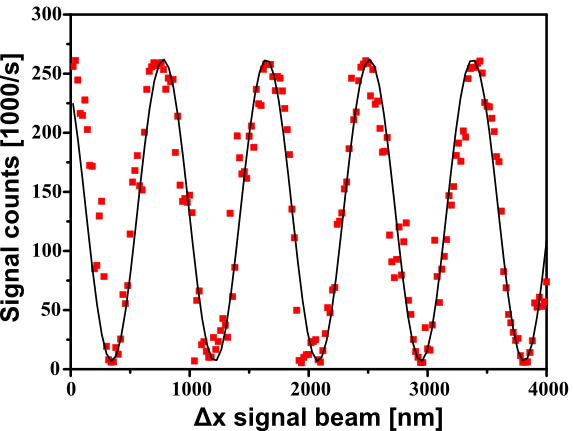}
\caption{Single-photon interference observed in counting signal photons at detector A when only BBO1 and BBO2 were pumped. The fringe visibility is 95\%.}
\label{Fig2}
\end{figure}

The observed single-photon interference at detector A of Fig. \ref{Fig1} as a function of the delay of the signal field s1 is shown in Fig. \ref{Fig2}. The fringe visibility in this measurement is very high, $V$ = 95\%, and larger than in previous spontaneous down-conversion, induced coherence experiments \cite{Wang91,Zou93}. The observed fringe spacing is given by the signal wavelength. The stimulated count rate in Fig. \ref{Fig2} is quite high as a consequence of the large average photon number of the stimulating He-Ne laser, and of course much higher than the count rate obtained in spontaneous down-conversion with the same pump power.

\subsection{BBO2 and BBO3 pumped}
\begin{figure}[t]
\includegraphics[width=7cm]{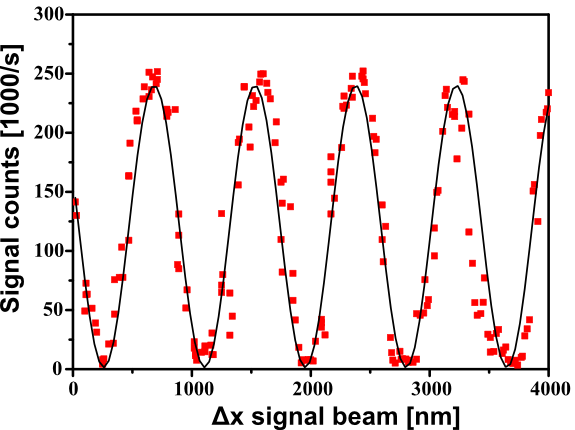}
\caption{Single-photon interference of signal photons counted at detector A when only BBO2 and BBO3 are pumped . The fringe visibility is 98\%.}
\label{Fig3}
\end{figure}

In the setup just discussed biphotons are also generated by {\sl spontaneous} down-conversion, which also
results in single-photon signal interference as a result of induced coherence \cite{heuer14-2}; this setup does not
distinguish between the interference due to spontaneous and stimulated coherence. However, two crystals can also
be synchronized in a ``parallel" arrangement when only BBO2 and BBO3 are pumped. In this case no induced coherence is expected 
because the idler modes i2 and i3 are distinct. The fields from the He-Ne laser are aligned to follow the same paths as the fields in modes i2 and i3 that would be spontaneously emitted by the two crystals near the He-Ne laser wavelength. The direction of these fields is determined by the position of detector A for the corresponding signal modes s2 and s3, which were again overlaid at a beam splitter. By moving the mirror (phase 1), single-photon interference fringes were obtained at detector A as shown in Fig. \ref{Fig3}. The fringe visibility ($V$=98\%) in this measurement is even higher than that of Fig. \ref{Fig2}.

\subsection{BBO1 and BBO3 pumped}
\begin{figure}[t]
\includegraphics[width=7cm]{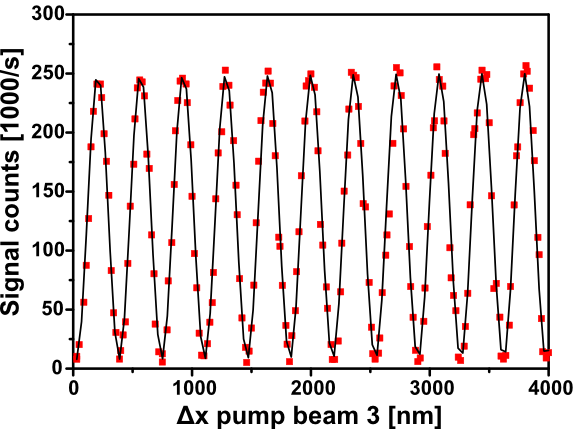}
\caption{Single-photon interference at detector A of signal photons generated in BBO1 and BBO3 as a function of the delay of the pump beam 3. The fringe visibility is 94\%.}
\label{Fig4}
\end{figure}

High-visibility signal interference fringes can also be obtained when two crystals BBO1 and BBO3 are in a cascade arrangement such that the signal modes s1 and s3 are as indistinguishable as possible and one pump beam is delayed with respect to the other. Without the He-Ne laser, induced coherence of the idler photons from BBO1 and BBO3 in this setup would be expected via the signal-photon channel. As previously shown \cite{heuer14-1}, the signal fields in modes s1 and s3 would not be in a fixed phase relation in that case and no interference between these channels is obtained. But by applying the He-Ne laser radiation to stimulate the down-conversion, coherence and single-photon interference can be realized between the two signal fields s1 and s3 from the two crystals by varying the delay between the two pump beams (delay in pump 3). The result of this measurement is shown in Fig. \ref{Fig4}. Again the fringe visibility is quite high ($V$ = 94\%) and the fringe spacing as expected is 355 nm, the pump wavelength. The interference in this case can be interpreted as a phase memory effect \cite{heuer14-2}.

These experiments demonstrate that stimulated coherence mainly just ``over-writes" any implications of induced coherence, but of course there are fundamental differences between the two effects. Note first that single-photon signal interference from induced coherence and spontaneous down-conversion (SPDC) is not observed in the experiments corresponding to Figs. \ref{Fig3} and \ref{Fig4}; spontaneously generated photon pairs appear only as a very small background signal that decreases the fringe visibility. This can be interpreted as a consequence of the fact that ``which-way" information about a signal photon at detector A in spontaneous down-conversion is made possible by the fact that its idler partner photon will be emitted from one or the other of the two crystals, and this specifies which of these crystals is the source of the signal photon. However, the rate of spontaneously generated signal photon counts is comparatively very small and does not significantly diminish the observed visibility determined by the stimulated down-conversion. (In all these experiments  the rate of spontaneously generated signal photon counts was only about 750 photons/sec.) The fringe visibility in the experiment corresponding to Fig. \ref{Fig4}, for example, is very high because the which-way information becomes effectively inaccessible due to the large and approximately equal numbers of idler photons in modes i1 and i3 when the stimulating He-Ne laser is applied.

\subsection{Three-crystal interference}

\begin{figure}[t]
\includegraphics[width=7cm]{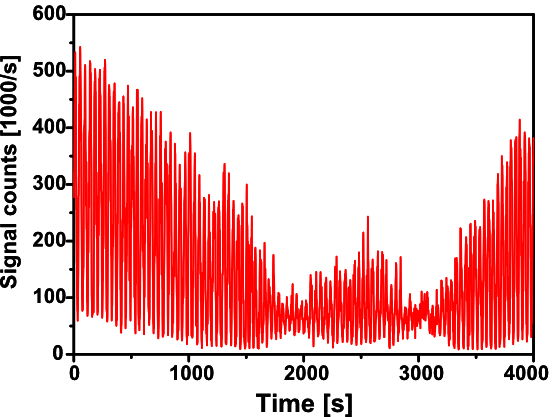}
\caption{Single-photon interference of signal photons generated in all three crystals as a function of time when the delays of the signal and the pump 3 fields were varied simultanously at a constant velocity. The fringe visibility is 94\%.}
\label{Fig5}
\end{figure}

In another experiment all three pump channels were opened. The delay lines for the signal and pump 3 fields were continuously moved with a constant velocity of 20 nm/s. The result of this single-photon interference measurement as a function of time is shown in Fig. \ref{Fig5}. Due to the simultaneous variation of two phase factors a strong beating in the detected signal was observed. The result looks quite similar to Fig. 5 of 
 Reference \cite{heuer14-1}, but here the observed interference is first-order whereas in the induced coherence measurement of Reference \cite{heuer14-1} interference was only observable in coincidence. Again the vacuum-field effects are ``over-written'' here.

\subsection{Variable transmission filter between BBO1 and BBO2}
\begin{figure}[t]
\includegraphics[width=8cm]{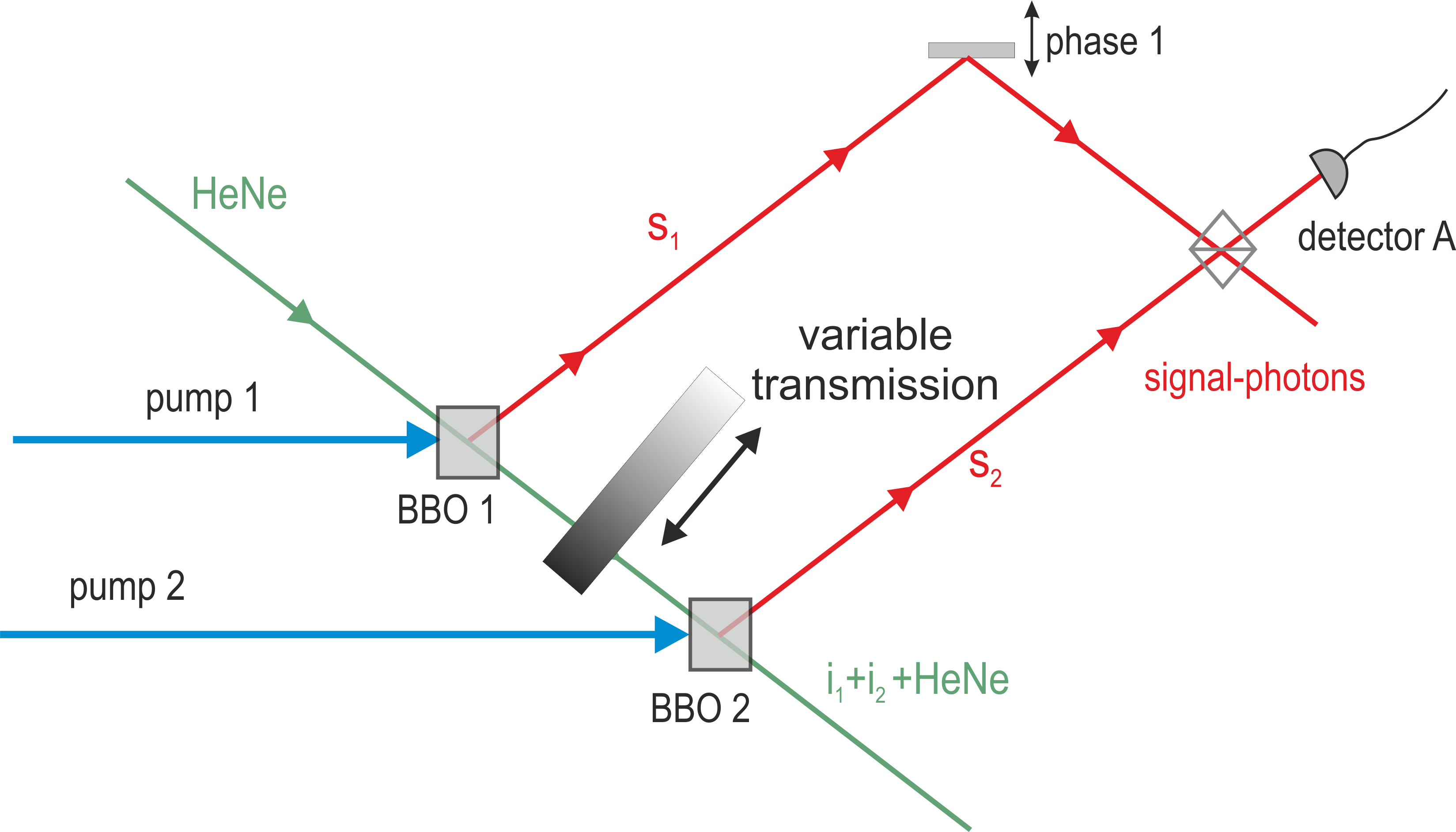}
\caption{Setup with two pumped crystals BBO1 and BBO2, with a variable filter placed between them.}
\label{Fig6}
\end{figure}

\begin{figure}[t]
\includegraphics[width=7cm]{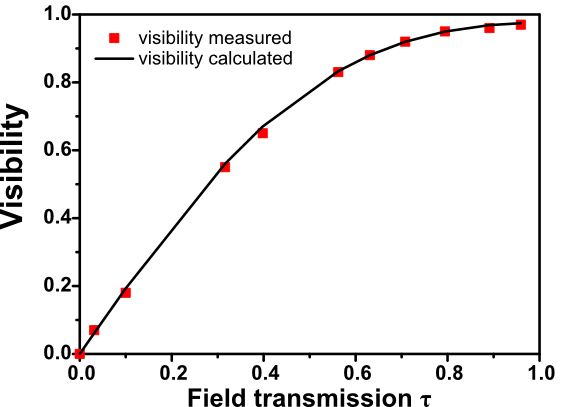}
\caption{Fringe visibility measured with the scheme of Fig. \ref{Fig6} as a function of the filter amplitude transmission $\tau$. The solid line is the calculated visibility from equation (\ref{Veq}).}
\label{Fig7}
\end{figure}

A more subtle difference between stimulated and induced coherence appears when a neutral density filter is introduced between the two crystals BBO1 and BBO2 of Fig. \ref{Fig1}. This arrangement is shown in Fig. \ref{Fig6}. The fringe visibility was measured in this setup by varying the phase delay (phase 1) and detecting the signal with the detector A, but with different filter amplitude transmissions $\tau$. The results of these measurements with the applied He-Ne laser as shown
in Fig. \ref{Fig6} are shown in Fig. \ref{Fig7}. The observed experimental data could be fit very well by the formula

\begin{equation}
\begin{aligned}
V = \frac{2 |\tau|}{1 + |\tau|^2}.
\end{aligned}
\label{Veq}
\end{equation} \\

\begin{figure}[t]
\includegraphics[width=7cm]{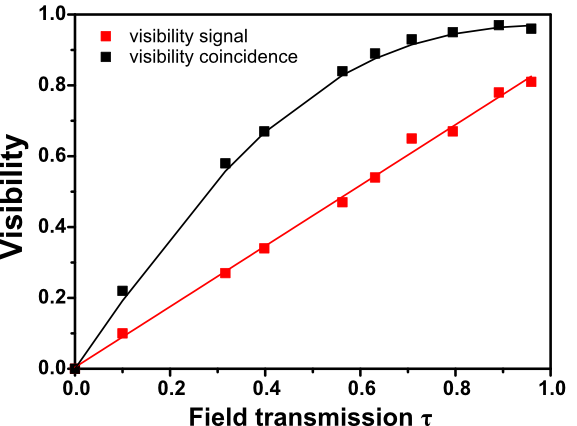}
\caption{Visibility vs. field transmission $\tau$ measured for single-photon signal interference (red curve) in the setup of Fig. \ref{Fig6}, but without the He-Ne laser. The black curve shows the measured visibility when signal and idler photons are counted in coincidence with two photodetectors.}
\label{Fig8}
\end{figure}

Without the He-Ne laser in Fig. \ref{Fig6}, i.e., in the case of single-photon interference via induced coherence, the
observed visibility is found to vary linearly with $|\tau|$, as shown in Fig. \ref{Fig8}. When signal and idler photons are counted in coincidence with two photodetectors, however, the observed visibility vs. $\tau$ shown in Fig. \ref{Fig8} is again found to follow equation (\ref{Veq}).

\section {Theory}
We can account for all these experimental results using a simplified theoretical model based on single-mode field operators in the Heisenberg picture \cite{remark1}. The simplification is made realistic by the approximate monochromaticity of the different fields and the mode alignments in the experiments. We start from an effective Hamiltonian in which the coupling of the pump (p), signal (s), and idler (i) fields is proportional to $a_p\ad_s\ad_i$ and its Hermitian conjugate, where as usual the $a$'s and $\ad$'s are photon annihilation and creation operators, respectively. The positive-frequency, photon annihilation parts of the signal and idler electric field operators are denoted by $\Ep_s$ and $\Ep_i$, respectively, and the negative-frequency, photon creation parts of these fields are similarly denoted by $\Em_s$ and $\Em_i$. The pump and He-Ne laser fields will be treated as undepleted, coherent fields, and we will assume that the crystals are lossless and oriented for perfect type I phase matching. 

Consider first the experimental setup shown in Fig. \ref{Fig1} when only BBO1 and BBO2 are pumped. The positive-frequency part of the signal electric field generated in crystal BBO1 and incident on detector A is expressed in our model as
\be
\Ep_{s1A}=iC_1\ad_{i1}e^{i\phi},
\label{th1}
\ee
where $\ad_{i1}$ is the photon creation operator for the idler field incident on BBO1. The constant $C_1$ is proportional to the nonlinear susceptibility of BBO1 as well as to the amplitude of the pump field incident on BBO1. The factor $ie^{i\phi}$ accounts for the phase accrued in propagation of the signal field to the mirror indicated in Fig. \ref{Fig1}, the reflection off the mirror, and the propagation to detector A. The positive-frequency part of the signal electric field generated in crystal BBO2 and incident on detector A is expressed similarly as
\be
\Ep_{s2A}=C_2\ad_{i2}.
\label{th2}
\ee
$\ad_{i2}$ is the photon creation operator for the idler field incident on BBO2 and $C_2$ is proportional to the nonlinear susceptibility of BBO2 and the amplitude of the pump field incident on BBO2.
We do not include explicitly any phase change of this signal field due to its propagation to A, so that $\phi$ in equation
(\ref{th1}) actually denotes the relative phase change in propagation of the two signal fields incident on A. We also do not
explicitly indicate the temporal variations $\exp(\pm i\om_st)$, $\exp(\pm i\om_it)$, and $\exp(\pm i\om_pt)$, as these variations have no effect on the calculated photon counting rates when we identify $\om_p=\om_s+\om_i$. The positive-frequency part of the total signal field at detector A (assuming that the beam splitter ratio is also contained in the constants $C_1$ and $C_2$) when BBO1 and BBO2 are pumped is 
\be
\Ep_{sA}=\Ep_{s1A}+\Ep_{s2A}=iC_1\ad_{i1}e^{i\phi}+C_2\ad_{i2},
\label{th3}
\ee
and the negative-frequency part of this field is
\be
\Em_{sA}=-iC_1^*a_{i1}e^{-i\phi}+C_2^*a_{i2}.
\label{th4}
\ee 
Treating detector A as a perfectly efficient, broadband photodetector, we take the signal photon count rate at A to be proportional to the normal-ordered expectation value
\bea
R&=&\la \Em_{sA}\Ep_{sA}\ra=|C|^2\big[\la a_{i1}\ad_{i1}\ra+\la a_{i2}\ad_{i2}\ra \nonumber\\
&&\mbox{}+ie^{i\phi}\la a_{i2}\ad_{i1}\ra -ie^{-i\phi}\la a_{i1}\ad_{i2}\ra\big],
\label{th5}
\eea
where for simplicity we take $C_1=C_2\equiv C$, consistent with the approximately equal powers of the pump fields incident on BBO1 and BBO2. As in Reference \cite{heuer14-1} we are assuming throughout that the down-conversion efficiency is small and include only terms to lowest order
in $C$.

Now in the experiment in which only BBO1 and BBO2 are pumped the idler modes for the two crystals may be assumed to be identical. This is the origin of the induced coherence between the signal fields from BBO1 and BBO2 when the idler modes for the two crystals are aligned and there is no externally applied idler field corresponding to the He-Ne laser field in Fig. \ref{Fig1}. Similarly, assuming the He-Ne laser field as well as the two pump fields (pump 1 and pump 2) in Fig. \ref{Fig1} are essentially undepleted in the down-conversion, the same laser field {\sl stimulates} the generation and coherence of the signal fields from the two crystals. Then, taking i1 and i2 to be identical in equation (\ref{th5}) \cite{remark},
\bea
R&=&2|C|^2\big(1-\sin\phi)\la a_{i1}\ad_{i1}\ra\nonumber\\
&=&2|C|^2\big(1-\sin\phi\big)\big[\la\ad_{i1}a_{i1}\ra+1\big]\nonumber\\
&=&2|C|^2\big(1-\sin\phi\big)\big[|\alpha_L|^2+1\big]\nonumber\\
&\cong&2|C|^2(1-\sin\phi)|\alpha_L|^2
\label{th6}
\eea
if the He-Ne laser field is described by a coherent state $|\alpha_L\ra$ ($a_{i1}|\alpha_L\ra=\alpha_L|\alpha_L\ra$), or simply
taken to be a classical field with a complex amplitude proportional to $\alpha_L$, $|\alpha_L|^2\gg 1$.
Since the phase $\phi$ comes from signal field propagation, $R$ varies sinusoidally with a period equal to the
signal field wavelength, as seen in Fig. \ref{Fig2}. The term $\la\ad_{i1}a_{i1}\ra$ accounts for the {\sl stimulated} generation  of the signal fields from the two crystals. If there is no applied idler field incident on BBO1, $\la\ad_{i1}a_{i1}\ra=0$ and $R$ then gives the interference pattern due to induced coherence \cite{Wang91,heuer14-2,heuer14-1}. Equation (\ref{th6}), based on the assumption of coherent applied fields, predicts
a fringe visibility of unity, consistent with the nearly perfect visibility observed (Fig. \ref{Fig2}).

To account as simply as possible for the single-photon signal interference observed in the experiment in which only BBO2 and BBO3 are pumped, let $a_{i1}\rightarrow a_{i3}\equiv a_{i30}+\alpha_L$ and $a_{i2}=a_{i20}+\alpha_L$ in Eq. (\ref{th5}), where $a_{i30}$ and $a_{i20}$
are annihilation operators for the vacuum idler modes incident on BBO3 and BBO2, respectively, and $\alpha_L$ is the complex amplitude describing the He-Ne laser fields incident on BBO3 and BBO2, again treating these fields as perfectly coherent and identical classical fields. In this experiment there is no induced coherence between the generated signal fields, since the vacuum fields at the two crystals are associated with distinct modes in this parallel arrangement. Therefore $\la a_{i30}\ad_{i20}\ra=\la a_{i20}\ad_{i30}\ra=0$ and, from (\ref{th5}) with $1\rightarrow 3$,
\bea
R&=&|C|^2\big[\la a_{i30}\ad_{i30}\ra+|\alpha_L|^2+\la a_{i20}\ad_{i20}\ra+|\alpha_L|^2\nonumber\\
&&\mbox{}+ie^{i\phi}|\alpha_L|^2-ie^{-i\phi}|\alpha_L|^2\big]\nonumber\\
&=&2|C|^2\big[1+|\alpha_L|^2(1-\sin\phi)\big]\nonumber\\
&\cong&2|C|^2(1-\sin\phi)|\alpha_L|^2,
\label{th7}
\eea
as in equation (\ref{th6}), where now we have taken $\la a_{i3}\ad_{i2}\ra\cong\la a_{i2}\ad_{i3}\ra\cong |\alpha_L|^2\gg 1$. The fact that there is induced coherence when only BBO1 and BBO2 are pumped, but only stimulated coherence when only BBO2 and BBO3 are pumped, does not affect the observed signal-photon count rates because the stimulated down-conversion is much stronger than the spontaneous down-conversion.

When only BBO1 and BBO3 are pumped, similarly, 
\be
E_{sA}^{(+)}=\big[C_1\ad_{i1}+C_3e^{i\phi_p}\ad_{i3}\big]ie^{i\phi},
\label{th8}
\ee
where $\phi$ is again the relative phase due to propagation of the two fields at A and $\phi_p$ is the phase delay of pump 3 relative to pump 1. The photon count rate at A in our model is then proportional to
\be
R=|C|^2\Big[\la a_{i1}\ad_{i1}\ra+\la a_{i3}\ad_{i3}\ra+\la a_{i1}\ad_{i3}\ra e^{i\phi_p}+\la a_{i3}\ad_{i1}\ra e^{-i\phi_p}\Big]
\label{th9}
\ee
for $C=C_1\cong C_3$. Again assuming that the stimulated down-conversion due to the He-Ne laser field is much stronger than
the spontaneous down-conversion, we take $\la a_{i1}\ad_{i1}\ra\cong\la a_{i3}\ad_{i3}\ra\cong|\alpha_L|^2$ and 
$\la a_{i1}\ad_{i3}\ra\cong \la a_{i3}\ad_{i1}\ra\cong |\alpha_L|^2$, assuming these fields have the same intensity. Then
\be
R\cong 2|C|^2[1+\cos\phi_p]|\alpha_L|^2.
\label{th10}
\ee
The interference in this case varies sinusoidally with the {\sl pump} wavelength (355 nm), as seen in the data shown in 
Fig. \ref{Fig4}.

In the experiment in which all three crystals are pumped, the positive-frequency part of the signal electric field at detector A in our model is
\be
E_{sA}^{(+)}=C\big[\ad_{i2}+ie^{i\phi}\big(\ad_{i1}+\ad_{i3}e^{i\phi_p}\big)\big],
\ee
and the photon count rate at A, assuming $C= C_1\cong C_2\cong C_3$, is proportional to 
\bea
R&=&|C|^2\Big\{\big|1+ie^{i\phi}\big|^2\la a_{i1}\ad_{i1}\ra+\la a_{i3}\ad_{i3}\ra\nonumber\\
\label{threecrystal}
&+&ie^{i\phi_p}\big(e^{i\phi}-i\big)\la a_{i1}\ad_{i3}\ra-ie^{-i\phi_p}\big(e^{-i\phi}+i\big)\la a_{i3}\ad_{i1}\ra\Big\},\nonumber\\
\eea
where we have again taken the modes i1 and i2 to be identical. If there is no applied He-Ne laser field, $\la a_{i1}\ad_{i1}\ra=
\la a_{i10}\ad_{i10}\ra=1, \la a_{i3}\ad_{i3}\ra=\la a_{i30}\ad_{i30}\ra=1$, but $\la a_{i1}\ad_{i3}\ra=\la a_{i10}\ad_{i30}\ra=0$ since the modes i1 and i3 are distinct, and
\be
R=|C|^2\Big\{2\big(1-\sin\phi\big)+1\Big\}.
\ee
In this case BBO3 simply contributes an incoherent background, since the possible detection of an idler photon in mode i3 would imply that a signal photon at detector A must have had BBO3 as its source. Without this incoherent background there would be perfect fringe visibility, since the indistinguishability of the modes i1 and i2 makes it impossible to infer whether a signal photon counted at detector had BBO1 or BBO2 as its source.
With the stimulating He-Ne laser applied, we can approximate (\ref{threecrystal}) by
\bea
R=2|C|^2|\alpha_L|^2\Big\{1-\sin\phi-\sin(\phi+\phi_p)+\cos\phi_p\Big\},\nonumber\\
\label{threecrystal2}
\eea
where we have taken $\la a_{i1}\ad_{i3}\ra\cong\la a_{i3}\ad_{i1}\ra\cong |\alpha_L|^2$ and assumed that the mean stimulating photon numbers $\la a_{i1}\ad_{i1}\ra\cong \la a_{i3}\ad_{i3}\ra\cong|\alpha_L|^2\gg 1$. The approximation (\ref{threecrystal2}) is consistent with the data shown in Fig. \ref{Fig5} for the three-crystal experiment with $\phi$ and $\phi_p$ varied simultaneously.

Consider finally the experiment of Fig. \ref{Fig6}, which differs from that of Fig. \ref{Fig1} by the insertion of a filter between BBO1 and BBO2. The counting rate in our model is again proportional to $R$ as given by equation 
(\ref{th5}) and the approximations it is based on, but now we must account for the effect of the filter on the idler field incident on BBO2. This is straightforward in the stimulated down-conversion case where we approximate the stimulating field of the He-Ne laser as a prescribed classical field as above. In this approximation $a_{i1}$ is replaced by $\alpha_L$ and $a_{i2}$ by $\tau\alpha_L$, where $\tau=|\tau|e^{i\theta}$ is the field transmission coefficient of the filter. Then
\be
R=|C|^2|\alpha_L|^2\big[1+|\tau|^2-2|\tau|\sin(\phi+\theta)\big]
\label{th11}
\ee
and the visibility predicted by our model is given by equation (\ref{Veq}) (Fig. \ref{Fig7}):
\be
V=\frac{R_{\rm max}-R_{\rm min}}{R_{\rm max}-R_{\rm min}}=\frac{2|\tau|}{1+|\tau|^2}.
\label{th12}
\ee

In the induced coherence case (no stimulating He-Ne laser) we cannot simply take $a_{i20}=\tau a_{i10}$ as would be implied
by the classical field treatment leading to equation (\ref{th11}); this would violate the canonical commutation relation
$[a_{i20},\ad_{i20}]=1$. Instead we write
\be
a_{i20}=\tau a_{i10}+\mathcal{L},
\label{th13}
\ee 
where $\mathcal{L}$ is a Langevin ``noise" operator that commutes with $a_{i10}$, has zero expectation value for the initial vacuum idler state of interest  and satisfies the commutation relation $[\mathcal{L},\mathcal{L}^{\dag}]=1-|\tau|^2$, so that $[a_{i20},\ad_{i20}]=1$ 
as required \cite{boydshi}. Then, from (\ref{th5}), we have the vacuum expectation value
\bea
R&=&|C|^2\Big[\la \ad_{i10}a_{i10}+1\ra+\la\ad_{i20}a_{i20}+1\ra\nonumber\\
&&\mbox{}-ie^{i\phi}\la(\tau a_{i10}+\mathcal{L})\ad_{i10}\ra-
ie^{-i\phi}\la a_{i10}(\tau^*\ad_{i10}+\mathcal{L}^{\dag})\ra\Big]\nonumber\\
&=&2|C|^2\big[1-|\tau|\sin(\phi+\theta)\big],
\label{th14}
\eea
which implies the visibility
\be
V=|\tau|.
\label{th15}
\ee
Note that the filter in the setup of Fig. \ref{Fig6} does not affect $\la a_{i20}\ad_{i20}\ra$ ($=1$) but in effect
results in a vacuum-field correlation function 
\be
\la a_{i20}\ad_{i10}\ra=\tau.
\label{th16}
\ee

The rate for the coincidence counting of signal photons at the detector A of Fig. \ref{Fig6} and idler photons
at a detector D behind BBO2 is proportional in our model to ($C_1\cong C_2=C$)
\be
R_{SA,ID}=\la \Em_{SA}\Em_{ID}\Ep_{ID}\Ep_{SA}\ra,
\label{th17}
\ee
with
\bea
\Ep_{SA}&=&ia_{s10}e^{i\phi}+a_{s20}+C[ie^{i\phi}+\tau^*]\alpha_L \nonumber\\
&&\mbox{}+C[ie^{i\phi}\ad_{i10}+\ad_{i20}]
\label{th18}
\eea
and
\be
\Ep_{ID}=\tau\alpha_L+C[\ad_{s10}+\ad_{s20}].
\label{th19}
\ee
The coincidence counting rate in the purely stimulated case ($|\alpha_L|\gg 1$) is easily understood physically to be
the same as the signal-photon counting rate, and therefore to have a fringe visibility given by equation (\ref{th12}) when
$\phi$ is varied. In the induced coherence case ($\alpha_L=0$) we obtain from (\ref{th18}) and (\ref{th19}) the vacuum expectation value
\bea
R_{SA,ID}&=&|C|^2\Big\{\big\la a_{s10}\ad_{s10}a_{i10}\ad_{i10}\big\ra+\big\la a_{s20}\ad_{s20}a_{i20}\ad_{i20}\big\ra\nonumber\\
&&\mbox{}+\big\la a_{s20}\ad_{s20}a_{i10}\ad_{i10}\big\ra+\big\la a_{s10}\ad_{s10}a_{i20}\ad_{i20}\big\ra\Big\}\nonumber\\
&&\mbox{}+|C|^2\Big\{ie^{i\phi}\big\la\big[a_{s10}\ad_{s10}+a_{s20}\ad_{s20}\big]a_{i20}\ad_{i10}\big\ra\nonumber\\
&&\mbox{}-ie^{-i\phi}\big\la\big[a_{s10}\ad_{s10}+a_{s20}\ad_{s20}\big]a_{i10}\ad_{i20}\big\ra\Big\}\nonumber\\
&=&4+2ie^{i\phi}\la a_{i20}\ad_{i10}\ra-ie^{-i\phi}\la a_{i10}\ad_{i20}\ra.
\label{th20}
\eea
Then (\ref{th16}) implies
\be
R_{SA,ID}=4|C|^2\big[1-|\tau|\sin(\theta+\phi)\big]
\label{th21}
\ee
and a fringe visibility given by (\ref{th15}), in agreement with the data shown in Fig. \ref{Fig8}.

\section{Is There a Phase Relation between Signal and Idler Fields?}\label{sec:phase}
The calculations in the preceding section require no assumptions about relative phases of signal and idler fields. 
In fact different statements have been made regarding the question of whether there is a phase relation between the signal and idler fields generated in two-photon down-cnversion. For example,   Pe'er {\sl et al.} \cite{perev} derive the relation 
\be
\theta_s+\theta_i=\theta_p-\pi/2
\label{perev}
\ee
for the phases $\theta_s$, $\theta_i$, and $\theta_p$ of the signal, idler, and pump fields, respectively. 
Mandel {\sl et al.}, however, refer to ``the absence of a phase relation between the signal and idler waves," \cite{mandel86} and take the signal and idler phases ``to be random and uncorrelated" \cite{mandel88}. As we now discuss, these different statements about signal and idler phases apply in different limits, one where a ``phase locking" occurs in a nonlinear regime and
the other in a linear regime where the conversion efficiency is very low. 

The coupled equations describing the propagation of the positive-frequency parts of the signal, idler, and pump electric fields
are
\bea
\frac{\pa E_s^{(+)}}{\pa z}&=&-i\kappa E_p^{(+)}E_i^{(-)},\nonumber\\
\frac{\pa E_i^{(+)}}{\pa z}&=&-i\kappa E_p^{(+)}E_s^{(-)},\nonumber\\
\frac{\pa E_p^{(+)}}{\pa z}&=&-i\kappa E_s^{(+)}E_i^{(+)},
\label{coupledeqs}
\eea
where the coupling constant $\kappa$ is proportional to the nonlinear susceptibility. Suppose first that the pump, signal, and idler fields are sufficiently intense that they may be treated classically. In this case, following Pe'er {\sl et al.} \cite{perev}, we replace $E_j^{(+)}$ by $R_j\exp(i\theta_j)$, $j=s,i,p$, where the amplitudes $R_j$ and phases $\theta_j$ are classical variables. Then, from equations (\ref{coupledeqs}) follows, 
\bea
\frac{{\pa R}_s}{\pa z}&=&-\kappa R_iR_p\sin\Delta\theta,\nonumber\\
\frac{{\pa R}_i}{\pa z}&=&-\kappa R_sR_p\sin\Delta\theta,\nonumber\\
\frac{{\pa R}_p}{\pa z}&=&-\kappa R_sR_i\sin\Delta\theta,\nonumber\\
\frac{{\pa \Delta\theta}_p}{\pa z}&=&\kappa\cos\Delta\theta\Big[\frac{R_iR_p}{R_s}+\frac{R_sR_p}{R_i}-\frac{R_iR_s}{R_p}\Big],
\label{coupledeqs2}
\eea
with $\Delta\theta\equiv\theta_p-\theta_s-\theta_i$. As the amplitudes $R_i$ and $R_s$ increase with propagation,  $\cos\Delta\theta\rightarrow 0$ and we obtain the relation (\ref{perev}), as follows from the considerations of Pe'er {\sl et al.} or from direct numerical integration of equations (\ref{coupledeqs2}) with $R_s(0)$ and $R_i(0)$ assumed to be small compared to $R_p(0)$ and $\Delta\theta(0)$ chosen ``randomly."

If, however, we focus on spontaneous down-conversion such that the pump is essentially undepleted and treated approximately as a classical field, we can ignore the last of equations (\ref{coupledeqs}) and write
\bea
\frac{\pa E_s^{(+)}}{\pa z}&=&-i\chi E_i^{(-)},\nonumber\\
\frac{\pa E_i^{(+)}}{\pa z}&=&-i\chi E_s^{(-)},
\label{coup1}
\eea
where $\chi=\kappa A_p$, with $A_p$ the complex amplitude of the pump field. These linear equations can of course be solved exactly, but if the conversion efficiency over a propagation distance $L$ is very small we can approximate $E_s^{(+)}(L)$ and $E_i^{(+)}(L)$ by
\bea
E_s^{(+)}(L)&=&E_s^{(+)}(0)-iKE_i^{(-)}(0),\\
E_i^{(+)}(L)&=&E_i^{(+)}(0)-iKE_s^{(+)}(0),
\eea
where $K=\chi L$. It follows that $\la E_s^{(-)}(L)E_i^{(+)}(L)\ra=0$, where the expectation value refers to the initial condition at $z=0$ in which there are no signal or idler photons. If we interpret this correlation function as a measure of phase correlation, it follows that the signal and idler phases are uncorrelated, consistent with the remarks of Mandel {\sl et al.} \cite{mandel86,mandel88}. 

\section{Summary and Remarks}\label{sec:conclusions}

Here and in previous work \cite{heuer14-1} we have found spontaneous parametric down conversion to be useful for the investigation of the role of vacuum fields and which-path information in measurements relating to complementarity. Although the phases of the emitted signal and idler fields are generally random and uncorrelated as a consequence of the vacuum fields taking part in the down-conversion, experiments employing induced coherence allow in effect for the synchronization of the phases of single-photon signal fields if, for example, superposed idler modes of two crystals are aligned. The resulting high fringe visibility in first-order, single-photon interference is then attributable to the fact that the indistinguishability of the idler photon paths prohibits determination of which of the two crystals was the source of a signal photon. In terms of 
vacuum fields, the interference is a consequence of the fact that we may in effect take the vacuum expectation value $\la a_{i10}\ad_{i20}\ra=1$ when the idler modes i1 and i2 are identical [Eqs. (\ref{th5}) and (\ref{th6})].

Without induced coherence, e.g., when the idler modes i2 and i3 are distinct, $\la a_{i20}\ad_{i30}\ra=0$ and there is no first-order interference
of the signal photons in the experiments in which only BBO2 and BBO3 are pumped. In this case there is which-path information
available for a signal photon via the detection of its idler photon partner, regardless of whether the experiment is set up to actually count an idler photon. But applying a laser field in the idler photon modes of the separated crystals makes it is possible to ``over-write"� the randomness of the vacuum fields and realize very high visibilities in these single photon interference experiments, as shown here using up to three down-conversion crystals in sequential, parallel or triangle schemes. In these stimulated coherence experiments there are so many  photons occupying the idler modes i2 and i3, for example, that information is largely lost as to the path associated with a detected signal photon, and very high visibilities, 
above 95\%, could be observed. In the experiment corresponding to Fig. \ref{Fig3}, for example, the calculated visibility $V=n/(n+1)$, where $n=|\alpha_L|^2$ is the mean number of He-Ne photons incident on BBO2 and BBO3 [see Eq. (\ref{th7})], and only a few stimulating photons are needed to realize high fringe visibilities.  

When a filter with amplitude transmission $\tau$ is inserted between two sequential crystals with induced coherence, the visibility of the single-photon signal interference fringes is given by $V=|\tau|$, whereas the visibility when signal and idler photons are
counted in coincidence is found to be $V=2|\tau|/(1+|\tau|^2)$. These experimental results, which were previously obtained in related experiments
by Wang {\sl et al.} \cite{Wang91}, are explained by our simplified theoretical model, as are the corresponding results in the case of stimulated coherence, where the single-photon and coincidence visibilities are both given by $V=2|\tau|/(1+|\tau|^2)$. All the experimental results in
these experiments on induced and stimulated coherence are in fact explained by our very simple theoretical model.

Finally we briefly addressed the question of whether there is a phase relation [Eq. (\ref{perev})] between signal and idler fields in parametric down-conversion, and concluded that there is no such relation in the case of spontaneous down-conversion, whereas such a relation holds in
the nonlinear regime of stimulated down-conversion.



\begin{thebibliography}{99}
\bibitem{remark} We ignore throughout any phase difference between the He-Ne laser fields incident on different crystals, as it only shifts the variable
phases of interest (e.g. $\phi$) by a trivial constant.
\end{thebibliography}
\end{document}